\documentclass[aip,rsi,reprint,graphicx]{revtex4-1}
\usepackage{graphicx}
\usepackage[colorlinks,
            linkcolor=blue,
            anchorcolor=blue,
            citecolor=blue
]{hyperref}

\begin{document}

\title{Neutron penumbral imaging simulation and reconstruction for Inertial
Confinement Fusion Experiments}

\author{Xian-You Wang}
\email[]{wangphysics@126.com}
\address{Department of Physics, Chongqing University, Chongqing 400044, P.R.
China}
\affiliation{Institute of High Energy Physics, Chinese Academy of Sciences,
Beijing 100049, P.R. China}
\author{Zhen-yun Fang}
\address{Department of Physics, Chongqing University, Chongqing 400044, P.R.
China}
\author{Yun-qing Tang}
\email[]{tangyq@126.com}
\address{Department of Physics, Chongqing University, Chongqing 400044, P.R.
China}
\affiliation{Institute of theoretical  Physics, Chinese Academy of Sciences,
Beijing 100049,P.R. China}
\author{Zhi-Cheng Tang}
\author{Hong Xiao}
\author{Ming Xu}
\address{Institute of High Energy Physics, Chinese Academy of Sciences,
Beijing 100049, P.R. China}

\date{\today}

\begin{abstract}
Neutron penumbral imaging technique has been successfully used as the diagnosis
 method in Inertial Confined Fusion. To help the design of the imaging systems in the future
 in CHINA. We construct the Monte carlo imaging system by Geant4. Use the point spread function
from the simulation and decode algorithm (Lucy-Rechardson algorithm) we got the recovery image.
\end{abstract}

\maketitle

Inertial confinement fusion (ICF) has attracted
much attention from all over the world. The National Ignition
Facility (NIF) programs in America and Laser MegaJoule (LMJ)
 in France aim at reaching the inertial confinement fusion of a
deuterium-tritium (DT) filled cryogenic target. China's ICF program
are also in progress, this paper's aim is just study the neutron
diagnosis system which will be used in the future ICF program in China.

\begin{figure}
\begin{center}
\begin{minipage}[t]{0.2\textwidth}
\includegraphics[width=\textwidth]{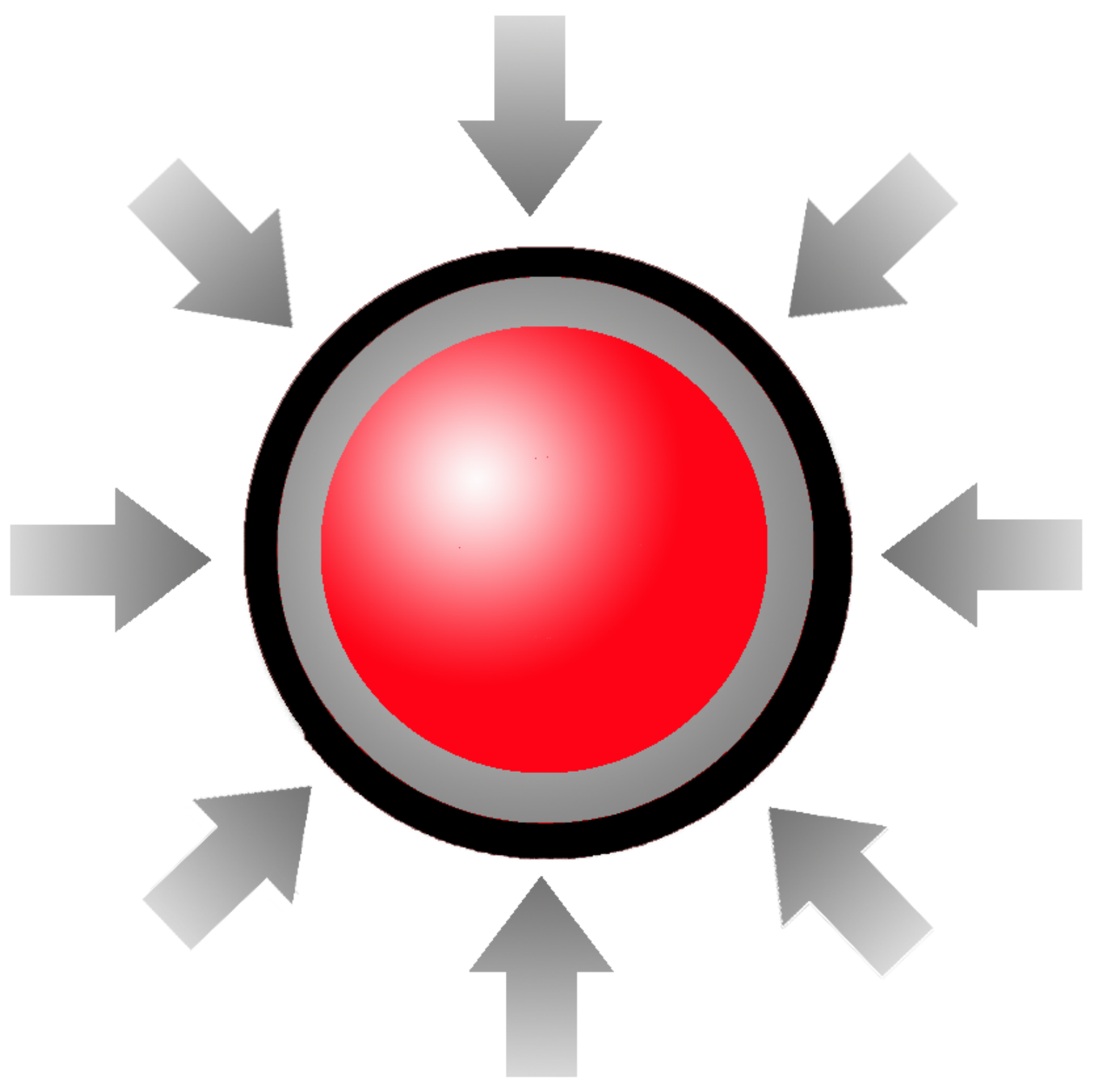}
\end{minipage}
\begin{minipage}[t]{0.2\textwidth}
\includegraphics[width=\textwidth]{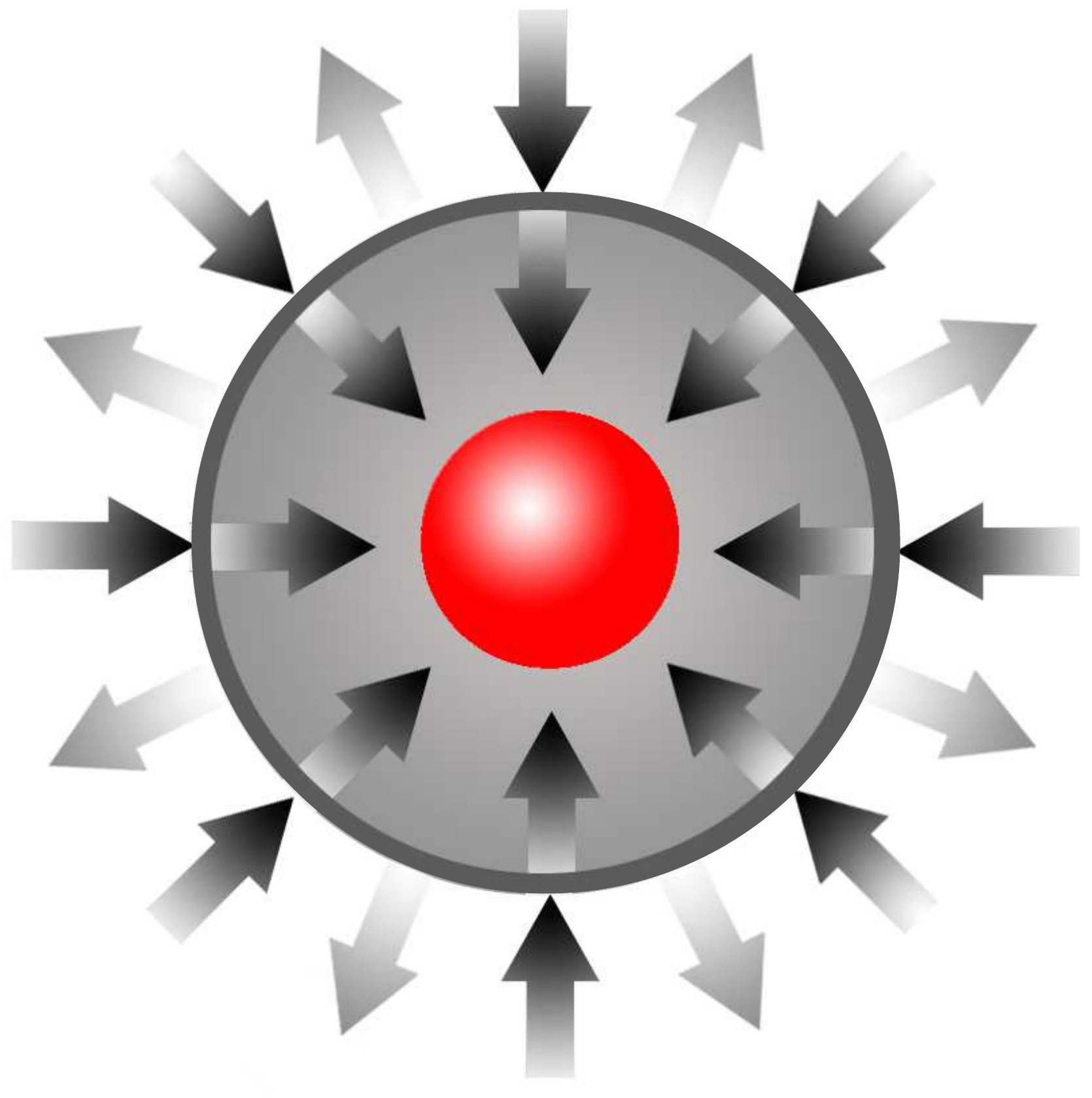}
\end{minipage}
\begin{minipage}[t]{0.2\textwidth}
\includegraphics[width=\textwidth]{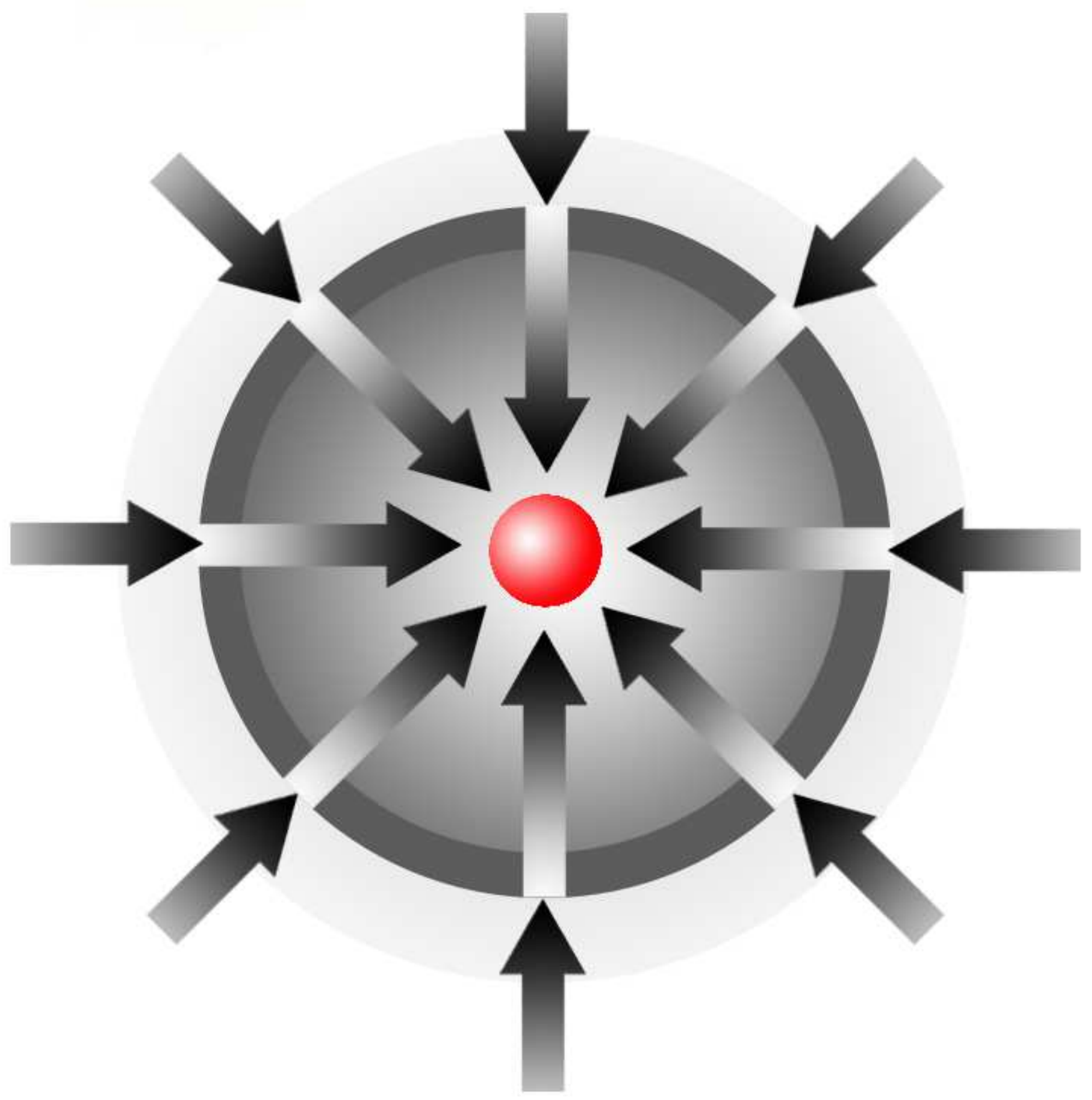}
\end{minipage}
\begin{minipage}[t]{0.2\textwidth}
\includegraphics[width=\textwidth]{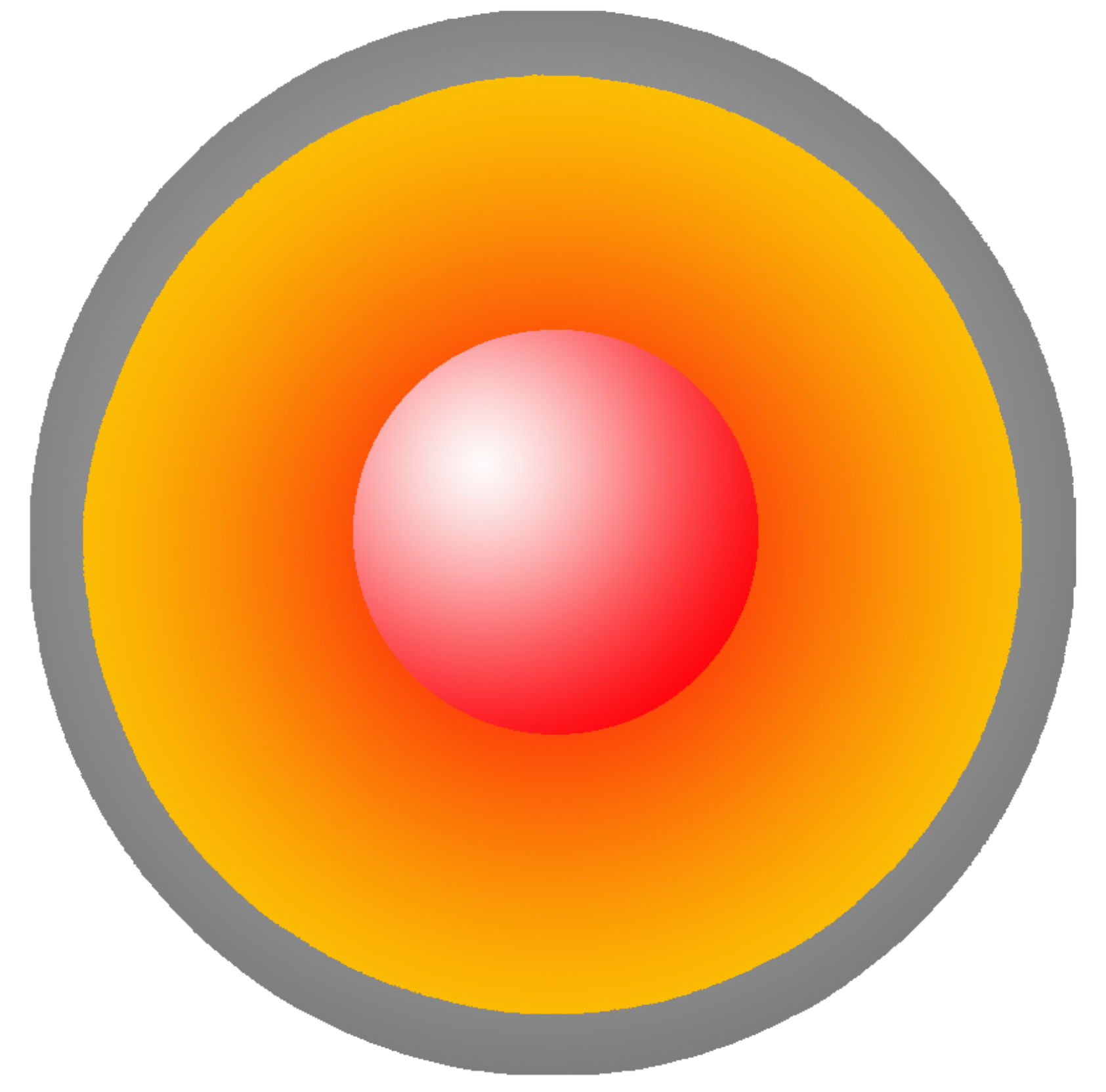}
\end{minipage}
\caption{ICF physics process (Radiation compressed, Implosion, Hot spot creation, Combustion)}{\label{fig:ICF}}
\end{center}
\end{figure}

The process of ICF can be divided into four steps as Fig \ref{fig:ICF}
: First step generate radiation to drive implosion.
 There are two approaches to drive implosion, one is laser
 direct driven, another is laser energy injected inside the
cavity  generates  X-rays that heat the DT fuel.  Under  X-ray
or laser compression,  the target density as well as  the mean
ions' temperature increases. Fusion reactions will occur within a hot
spot  generating 14 MeV neutrons and alpha particles\cite{ICFSim,ICFSim1}.
If the density is sufficient to trap the energetic alphas or other
radiation, these energy will be deposited on the surrounding  cold  DT
medium. Then the chained thermonuclear reactions will lead to a
combustion in the compressed target. For a well compressed target,
 combustion will last long enough to generate more energy than the injected energy.
 If ever, drive symmetry is poor or pulse shaping is imprecise,
 burning will be low. So we can use the neutrons escaped
 from the compressed cores of target to study the compressed process
 to make sure the drive be symmetrical.

There are several neutron imaging techniques for inertial confinement
fusion (ICF) experiment to detect the size, shape, and uniformity of the
compressed cores. The pinhole imaging is the easiest way, but the low
neutron yield has an effect on the sensitivity. In NIF experiment a
pinhole array is designed to increase the sensitivity\cite{pinhole1,
pinhole2,pinhole3,pinhole4,pinhole5}.

The another technique is the neutron penumbral imaging (NPI), which have been
widely studied by LMJ\cite{penum1,penum2,penum3,penum4,penum5,penum6} and other groups
\cite{penum_other1,penum_other2,penum_other3,penum_other4,penum_other5,penum_other6}.
Compared with other neutron imaging techniques, the sensitivity of neutron
penumbral imaging is higher. Annular imaging is nearly the same as penumbral
imaging, the only difference of the imaging process is aperture.

In this paper we just based on the simulation model in the reference\cite{cqu1,cqu2,cqu3,cqu4}
to design a software package NGPINEX which can be used to simulate the
neutron imaging process and recovery the coded image.
The imaging process simulation have been described in Geant4,
and then the recover algorithm we used Lucy-Richardson method\cite{LR1,LR2},
 which have been successfully applied on decoding  x-ray ring code image\cite{xrayLR}.
 The whole program is written in C++, which can be
run under the linux system with multi-core server.

In the program we simulated neutron penumbral imaging and Annular imaging,
the pinhole imaging block is also included in the program. Based on the NIF and LMJ experiment
we design an imaging system aimming to achieve $20\mu m$ resolution.
At last we use the Lucy-Richardson method successfully got the
recovered image with $20\mu m$ resolution

\section{The imaging system simulation}
\begin{figure}
\begin{center}
\includegraphics[scale=0.75]{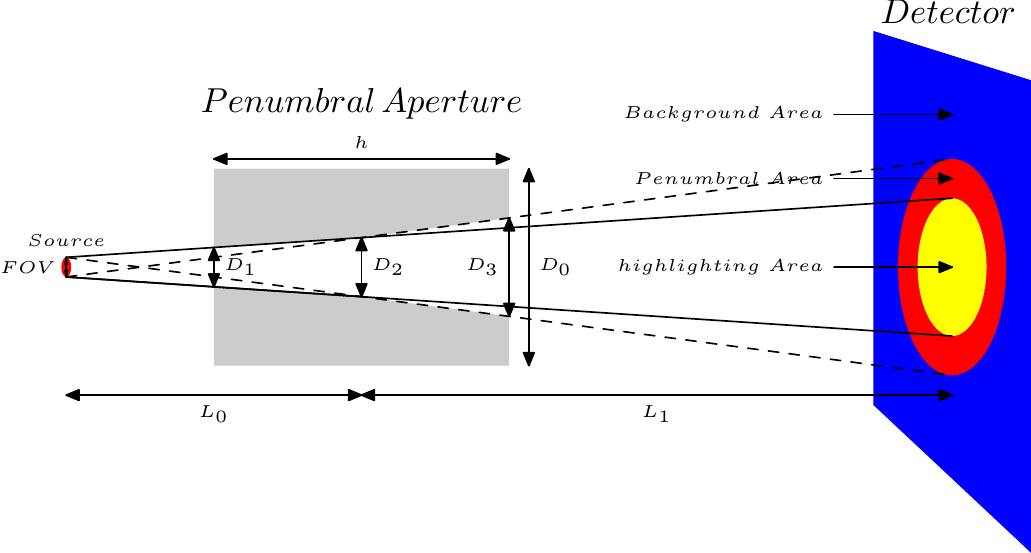}
\caption{Penumbral imaging}{\label{fig:penumbral}}
\end{center}
\end{figure}

The structure of neutron penumbral imaging system
is shown in Fig.\ref{fig:penumbral}. The source
, aperture and detector composed the whole imaging system.
The fast neutrons emitting from a burning target are
scattered by the aperture materials, and form an
penumbral image on the detector arrays. In our simulation
the coded aperture is placed at $55 mm$ from the source, the
detector is  placed at $7m$ from the aperture.
The aperture is designed as a $50 mm(h)$ thick $10 mm$ diameter(tungsten) cylinder
 to achieve an effective absorption of the  $14 MeV$ neutrons  in the  opaque
parts  of the aperture. The pattern of the aperture is a disk
(penumbra) which consists in a biconical hole, the diameters
 $(D_{1}, D_{2}, D_{3})$ of the biconical hole are $0.3mm, 0.38mm, 0.535mm$.
 An enlarged  coded image is projected on the neutron detector
  which is placed on the target bay floor, at $7895 mm$
from  the aperture. The image obtained from the detector is a
convolution of the source spatial distribution with the aperture
transfer function.

In the Fig.\ref{fig:penumbral} only the red annular of the coded image contains
the effective information of the source shape changing. To achieve the high SNR
we can place a plug inside the hole of an aperture of the same dimensions as
the penumbra(Fig.\ref{fig:penumbral}) to define a annular, which will be study in
another paper. This pattern can shield a lot of background but also face the
alignment tolerance problem and not enough neutron can be detected.

\begin{figure}
\begin{center}
\begin{minipage}[t]{0.2\textwidth}
\includegraphics[width=\textwidth]{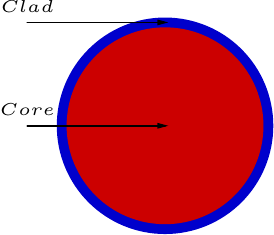}
\end{minipage}
\hfill
\begin{minipage}[t]{0.2\textwidth}
\includegraphics[width=\textwidth]{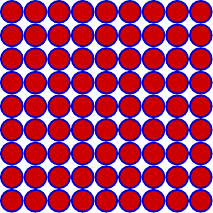}
\end{minipage}
\caption{Fiber array }{\label{fig:fiberArraySection}}
\end{center}
\end{figure}

The neutron detector is composed of $255 \times 255$ array of
scintillating fibers Fig.\ref{fig:fiberArraySection}. Each scintillating
fiber can be realized as a sub-detector, the structure was designed as
the left figure of Fig.\ref{fig:fiberArraySection}. The core part is a
tubes of $460 \mu m$ diameter and the material is polystyrene BCF-10
(molecular formula C6H5CH=CH2, $1.05g/cm^3$). 14 MeV neutrons mainly interact by elastic
scattering on hydrogen nuclei inside the fibers. As they lose
kinetic energy, the recoil protons will produce light, a part of the light
is guided through these fibers. Outside the core part is a layer of the clad
material, which is used to avoid the scintillating light interaction in
the array of scintillating fibers. The material of the envelope is Polyvinyl
alcohol(molecular formula CH2=CHOH, $1.26g/cm^3$) and the thickness is $20 \mu m$.
An optical relay then casts the scintillating light from the end of the array to an image
intensifier tube (IIT) which gates the 14 MeV neutrons at their
arrival time on the detector. The image is then reduced with a
fibered optical taper onto a CCD.
Overall spatial resolution of the system for both the annular
and the penumbral apertures is the same, which is determined by
the point spread functions (PSF) of the coded aperture and detector
resolution ($\Delta s_{detector}$) scaled down the source plane
as equation \ref{eq:resolution}. In the paraxial approximation,
the broadening of the apertures point spread function is identical,
determined by the FOV and the aperture to target distance ($L_0$).
 \begin{equation}
\label{eq:resolution}
\Delta s=\sqrt{\frac{ln(2)\times FOV}{2\times L_{0}\times\mu}
+\Delta s_{detector}^{2}\times\left(\frac{L_{0}}{L_{1}}\right)^{2}}
\end{equation}
where $\mu$ is the attenuation  of the neutrons in tungsten and $L_1$
is the detector to the target distance. With this design, the overall
spatial resolution of  the system is about 20 $\mu m$ in source plane.

In the GEANT4 simulation we used the physics list QGSP\_BERT\_HP
which contains the high precision neutron package (NeutronHP)
to simulate the process of neutrons' transportation with the energy
 below 20 MeV down to thermal energies.
The physical process neutron capture, fission, elastic
and inelastic scattering (including absorption) are treated
by referring to the ENDF-B VI cross-section data. Besides,
in the ref\cite{MCNP} has demonstrated the validity of GEANT4
calculations in neutron generation and transportation which
is the same as MCNP. Moreover, object-oriented programming and
highly sophisticated processing of both electromagnetic
interactions and ionization processes enhance GEANT4
capabilities for the study of mixed fields and complicated
geometries.

\begin{figure}
\begin{center}
\includegraphics[scale=0.75]{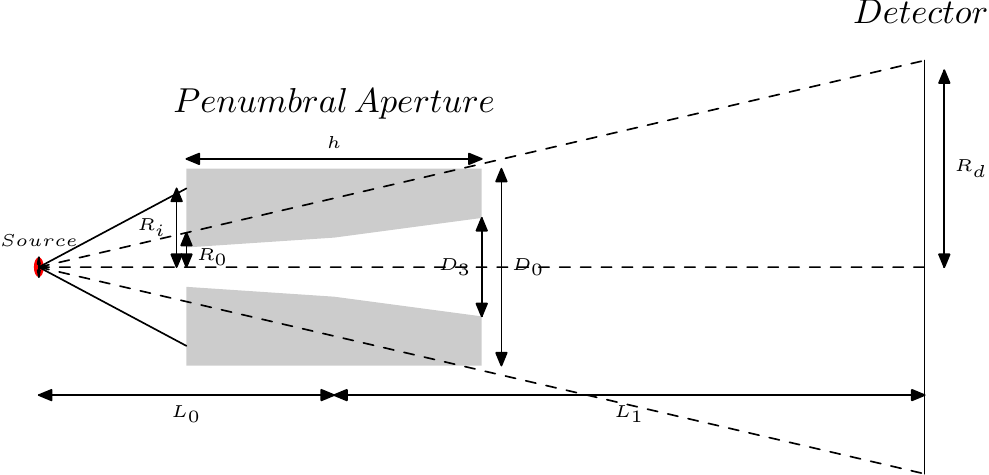}
\caption{The neutron radiated region in the simulation}{\label{fig:penumbral-simulate}}
\end{center}
\end{figure}

In the ICF experiment the neutrons generated by hot spot
spread in whole space. The idealized aperture only allow
the neutrons in the solid angle as Fig.\ref{fig:penumbral}
to deposit the energy on the detector. But actually because the
neutrons have the high ability to penetrate the start part of
the aperture, and also the neutrons will scatter with the wall of
the aperture, these neutrons which hit the detector will become
the background of the imaging. To simulate the complete imaging
process and also save the calculation time we ask the program
only generate random distribution neutron in the cone with $R_{i}$
radius as Fig.\ref{fig:penumbral-simulate}. As equation
\ref{eq:simulate-approximate3} $R_{i}$ is determined by $R_{0}$
which is the width of the detector project on the front face
of the aperture.

With this design there are about one half of the events be scattered by
the penumbral aperture.To get the high statistic sample we simulated
$1\times10^{10}$ effective events of penumbral imaging.

\begin{equation}
\label{eq:simulate-approximate3}
   \rm R_{i} = \left\{ \begin{array}{lll}
   D_{0}/2, & R_{0} > D_{0}/2 \\
   1.3R_{0}, &  D_{3}/2 \leq R_{0} \leq D_{0}/2 \\
   D_{3}/2, & R_{0} < D_{3}/2 \end{array} \right.
\end{equation}

\section{Reconstruction of the coded image}
At present, there are a lot of reconstruction methods
have been developed. Among them the simplest and quickest
method is Wiener filter method, which is a linear reconstruction
method. As we cannot get prior knowledge of the signal-noise ratio,
wiener filter method cannot get good results. Therefore we
can use this method as the reference or use the result as the
input of some nonlinear reconstruction method. In these years
there are some nonlinear reconstruct methods have been introduced to
reconstruct penumbral image such as genetic algorithm
and heuristic algorithm proposed by Chen et al.\cite{penum_other1},
and the classical molecular dynamics reconstruction
method proposed by Liu et al.\cite{penum_other2,penum_other3},

In the experiment L. Disdier et al. had successfully used the
filtered autocorrelation technique\cite{filauto1} to get
$20\mu m$ resolution reconstruction image from the penumbral
imaging system with $22.3\mu m$ theoritical resolution.
Filtered autocorrelation technique is different from other method
which don't need to rely on the simulation PSF in the reconstruction.

Another nonlinear method Lucy-Rechardson algorithm\cite{LR1,LR2} has been
successfully used on X-ray ring coded imaging in experiment\cite{xrayLR}.
And this algorithm has been most widely used for restoring images
with noise from counting statistics and the data approximate Poisson statistics.
In this program we first try to use the Lucy-Rechardson method to get reconstructed image.
The process of the imaging can be described as a spatial intensity distribution
$g(x,y)$ convolved by a neutron source spatial intensity distribution $f(x,y)$ with
a point spread function (PSF) $h(x,y)$. Add the noise $n(x,y)$ the detector
image can be expressed as
\begin{equation}
\label{eq-image}
g(x,y)=f(x,y)\otimes h(x,y)+n(x,y)
\end{equation}

Lucy-Rechardson algorithm is an iterative method. Based on the reference\cite{LR1,LR2},
the decode formulation can be expressed as \ref{eq-reco1}. Where PSF(h) and image on
detector(g) come from the simulation, the reconstructed image(f) can be extracted after k times
iterate.
\begin{equation}
\label{eq-reco1}
f_{k+1}=f_{k}(h\odot \frac{g}{h\otimes f_{k}})
\end{equation}

\begin{figure}
\begin{center}
\begin{minipage}[t]{0.22\textwidth}
\includegraphics[width=3.0cm]{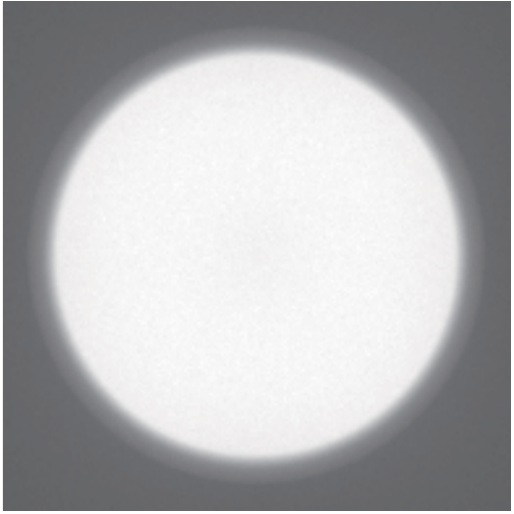}
\end{minipage}
\caption{Distribution of point spread function(PSF).}{\label{fig:psf-penumbral}}
\end{center}
\end{figure}

\begin{figure}
\begin{center}
\begin{minipage}[t]{0.22\textwidth}
\includegraphics[width=3cm]{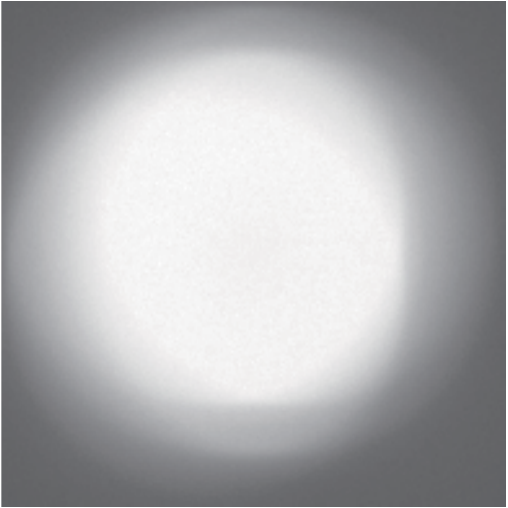}
\end{minipage}
\begin{minipage}[t]{0.22\textwidth}
\includegraphics[width=3cm]{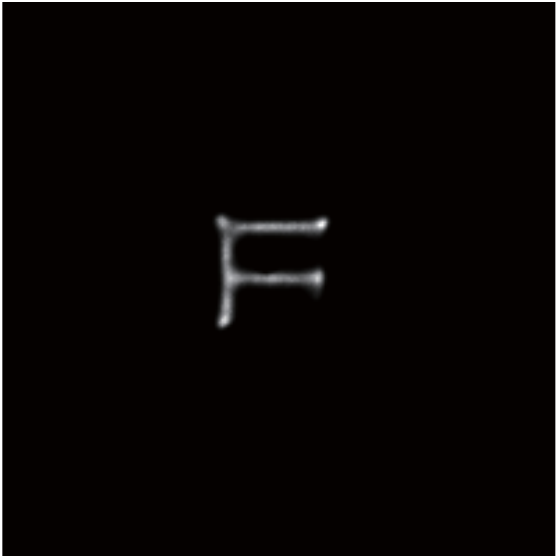}
\end{minipage}
\caption{`F' source imaging for penumbral aperture(left),
 unfolded result (right).}{\label{fig:image-f-penumbral}}
\end{center}
\end{figure}

The PSF for penumbral and annular imaging are presented in Fig. \ref{fig:psf-penumbral}
respectively. In Fig. \ref{fig:image-f-penumbral} we presented the raw images with a `F'
shape neutron source and the unfolded results with Lucy-Rechardson method. The width of the
`F' shape source is $20\mu m$.

\begin{figure}
\begin{center}
\begin{minipage}[t]{0.22\textwidth}
\includegraphics[width=3cm]{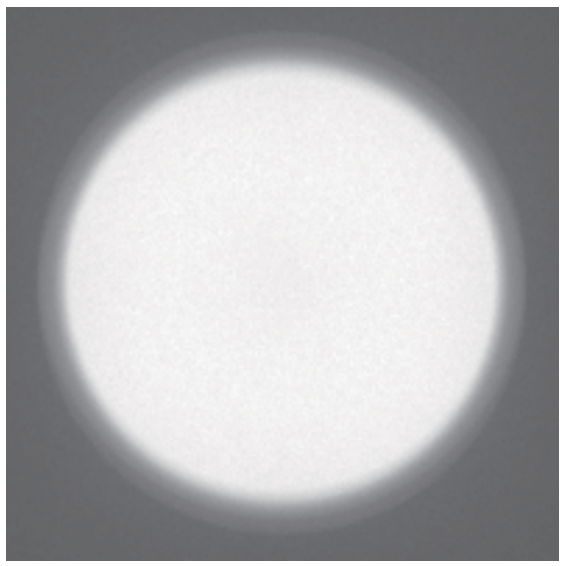}
\end{minipage}
\begin{minipage}[t]{0.22\textwidth}
\includegraphics[width=3cm]{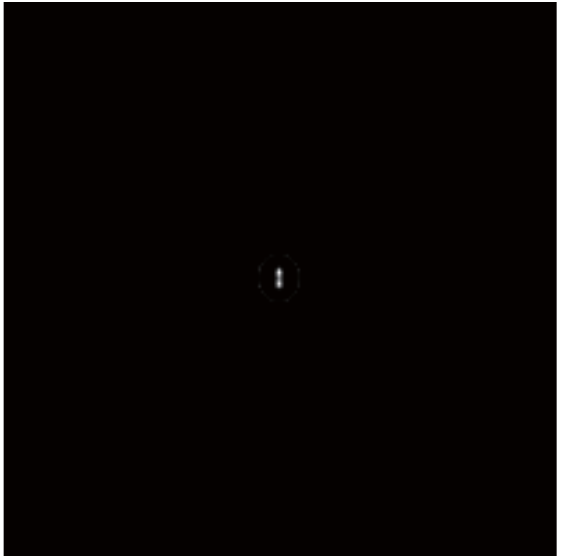}
\end{minipage}
\caption{Two point source imaging for penumbral aperture(left),
unfolded result(right).}{\label{fig:image-twopoint0.02mm-penumbral}}
\end{center}
\end{figure}

From the Fig. \ref{fig:image-f-penumbral} we can see with this design the imaging
system can achieve the aim to reconstruct the neutron source. With Lucy-Rechardson
method the unfolded image of Penumbral aperture is clear. To get
the real resolution of the system we simulate two point source with distance $20\mu m$
imaging on the detector.  In the Fig. \ref{fig:image-twopoint0.02mm-penumbral} present
the raw image and unfolded image.
For the annular imaging reconstruction and detail of the unfolding method
we will present in further study.

\section{Conclusion}
We have successfully construct one software NGPINEX which integrated
the imaging system simulation and the reconstruction method. The
simulation and reconstruction result show we can use this imaging system
with these parameters to get $20\mu m$ resolution image of DT implore.
To attend the aim of $5\mu m$ resolution in the future we can use
this software to optimize the system parameter.

\vspace{1cm}
{\bf Acknowledgments}:
 The authors would like to thank Ming Jiang and Bing-quan Hu
 for the help on the simulation model construction.
 This work was supported in part by the Fundamental Research
 Funds for the Central Universities under Grant No.CDJXS1102209
 and the Program for New Century Excellent Talents in Univresity
 under Grant No. NCET-10-0882, and by Natural Science Foundation
 of China under Grant No.10805082 and No.11075225.


\end{document}